\documentclass[twocolumn,showpacs,amsmath,amssymb,prl,superscriptaddress,floatfix]{revtex4-1}
\usepackage{bm}
\usepackage{amsmath,amsfonts}

\newcommand{\LL}{\mathcal{L}}
\newcommand{\I}{\mathbb{I}}

\begin{document}

\title{Comment on ``Abrupt transition in the structural formation of interconnected networks'', F.\ Radicchi and A.\ Arenas,     Nature Phys.\ {\bf 9}, 717 (2013)}

\author{Juan P. Garrahan}
\author{Igor Lesanovsky}
\affiliation{School of Physics and Astronomy, University of
Nottingham, Nottingham, NG7 2RD, UK}

\pacs{}

\begin{abstract}
A recent paper [F.\ Radicchi and A.\ Arenas, Nature Phys.\ {\bf 9}, 717 (2013)] presented the finding of an abrupt transition in the structure of interconnected networks.  This transition was said to be generic and to occur even in networks of finite size.  Furthermore, it was remarked that this singular behaviour could be understood in the spirit of a first-order phase transition.  We show here that the generic singularity found in that paper is a trivial consequence of the reducibility of the ``supra-Laplacian'' operator studied. The singular changes observed are therefore not related to any collective abrupt structural transformation in the interconnected networks.
\end{abstract}

\maketitle

Phase transitions are collective phenomena that occur in the limit of large system size \cite{books}.
For static phase transitions the relevant limit is that of volume or number of particles tending to infinity, while for dynamical transitions the large size limit can also include that of large time.
Often, phase transitions can be described by a singular change in the spectrum of a matrix (such as a transfer matrix for a partition sum, or a Hamiltonian for a quantum system). In this case the large size limit is one where the matrix dimension also tends to infinity. Technically this is important as singularities in the spectrum of irreducible matrices describing physical systems will typically only emerge when approaching this limit. Connected to this is for example the no-crossing theorem that holds for the eigenvalues of Hermitian matrices \cite{Neumann1929}. In general one must therefore be cautious when a singularity which occurs in a {\em finite} system---such as an eigenvalue crossing---is identified with a phase transition, since this is typically an indication of the reducibility of the matrix (often due to a symmetry), and not of a collective effect.  We show that, as expected from these generic arguments, the behaviour of the coupled networks of Ref.\ \cite{Radicci2013} is not an exception. The singularities interpreted there as phase transitions are really a trivial manifestation of the reducibility of the relevant operator, i.e., that it has a block structure.

This is directly seen by writing the ``supra-Laplacian'' of Eq.\ (2) of Ref.\ \cite{Radicci2013} as:
\begin{equation}
\LL = \frac{\LL_{A}+\LL_{B}}{2} \otimes \I + \frac{\LL_{A}-\LL_{B}}{2} \otimes \sigma_{z} + p~ \I \otimes (1-\sigma_{x}) ,
\nonumber
\end{equation}
where $\otimes$ indicates direct product and $\sigma_{x,z}$ are two-dimensional Pauli matrices. The Pauli matrices can be thought of as acting on a (global) degree of freedom which labels each of the layers of the network.  That is, layer A and B are represented, respectively, by the eigenstates $\left|\uparrow\right>$ or $\left|\downarrow\right>$, with eigenvalues $\pm 1$, of $\sigma_z$.

An immediate observation is that when $\LL_{A}=\LL_{B}$ (the first example discussed in \cite{Radicci2013}) the matrix $\LL$ is the sum of two operators that can be diagonalised independently, and hence $\LL$ is trivially reducible.  The lowest eigenvalues of $\LL_{A}$ are zero and $\lambda_{2}$, and those of $p(1-\sigma_{x})$ are zero and $2p$. The eigenvectors and eigenvalues $e_j$ of the lowest and the relevant excited states are thus given by
\begin{eqnarray*}
    \left|g\right>& \equiv &\left|1\right>\otimes \frac{\left|\uparrow\right>+\left|\downarrow\right>}{\sqrt{2}}, \quad e_g=0,\\
    \left|s_+\right>& \equiv &\left|2\right>\otimes \frac{\left|\uparrow\right>+\left|\downarrow\right>}{\sqrt{2}}, \quad e_+=\lambda_2, \\
  \left|s_-\right>& \equiv &\left|1\right>\otimes \frac{\left|\uparrow\right>-\left|\downarrow\right>}{\sqrt{2}},\quad e_-=2p
\end{eqnarray*}
where $\LL_A \left|1\right> = 0$ (ground state of $\LL_A$) and $\LL_A \left|2\right> = \lambda_2 \left|2\right>$ (first excited state of $\LL_A$). It follows immediately that the excited state will cross when $\lambda_{2}=2p$. The origin and nature of the ``radical change of the structural properties of the system''  \cite{Radicci2013} is just a consequence of the form of these eigenstate: For $p < p^*=\lambda_2/2$ the state $\left|s_-\right>$ is the first excited state and hence the entries of the eigenvectors that correspond to nodes of network A and B have the opposite relative sign. When $p > p^*=\lambda_2/2$ the first excited state is $\left|s_+\right>$ and the nodes have the same sign. Since $\left<1|2\right>=0$ and the entries of $\left|1\right>$ are equal, some components of $\left|2\right>$ within the individual network layers must have opposite signs. This is however a generic feature of the first excited state of a non-trivial Laplacian.  

The explanation above is equivalent to the observation that this problem actually maps to two disconnected networks.  In this latter system a trivial eigenvalue crossing can always be made to occur when the gap of the Laplacian of one of the independent networks is changed with respect to the other.  Clearly in such a system one would not consider the eigenvalue crossing as a phase transition.  
Furthermore, in such eigenvalue crossing there is no ``discontinuity in the first derivative of an energy-like function''  \cite{Radicci2013} (and therefore it cannot be interpreted as a first-order transition) given that each eigenvalue varies smoothly and the apparent singularity is due to not following the correct eigenvalue as the relevant parameter, $p$ in the case above, is varied. 

The situation does not change when $\LL_A \neq \LL_B$. The reason is that $\left|s_-\right>$ remains an exact eigenstate with an eigenvalue $2p$ (irrespective of the difference between $\LL_A$ and $\LL_B$). Thus $\LL$ remains reducible and a crossing in the second smallest eigenvalue can be achieved for a suitable value of $p$.  From this analysis of the spectrum of the supra-Laplacian it is evident that there are neither collective effects at work nor that one can regard the phenomenon observed in Ref. \cite{Radicci2013} as a phase transition.  The singular behaviour is merely a consequence of the reducibility of $\LL$.  Moreover, given that $\LL$ resembles the dynamic operator for state-dependent diffusion of a single spin-half particle on a network it appears unlikely that even in the limit of the network size going to infinity a true phase transition will emerge.  It is thus difficult see how to support the anticipated consequences, for example, that ``any real-world interconnected system is potentially at risk of abrupt changes in its structure'' \cite{Radicci2013}.


\thebibliography{9}

\bibitem{books} See for example,
N. Goldenfeld, Lectures On Phase Transitions And The Renormalization Group (Westview Press, Boulder, 1992); S. Sachdev, Quantum Phase Transitions (Cambridge University Press, 2011); L. Peliti, Statistical mechanics in a nutshell (Princeton University Press, 2011).


\bibitem{Radicci2013} F.\ Radicchi and A.\ Arenas, Nature Phys.\ {\bf 9}, 717 (2013).
\bibitem{Neumann1929} J. v. Neumann and E. Wigner, Phys. Zeitschr. XXX, 295 (1929).
\end{document}